# On the Violation of Honesty in Mobile Apps: Automated Detection and Categories

### Humphrey O. Obie
HumaniSE Lab
Monash University
Melbourne, Australia
humphrey.obie@monash.edu

### Idowu Ilekura
Data Science Nigeria
Lagos, Nigeria
ilekuraidowu@gmail.com

### Hung Du
Applied Artificial Intelligence Inst.
Deakin University
Melbourne, Australia
hung.du@deakin.edu.au

### Mojtaba Shahin
School of Computing Technologies
RMIT University
Melbourne, Australia
mojtaba.shahin@rmit.edu.au

### John Grundy
HumaniSE Lab
Monash University
Melbourne, Australia
john.grundy@monash.edu

### Li Li
Faculty of IT
Monash University
Melbourne, Australia
li.li@monash.edu

### Jon Whittle
CSIRO's Data61
Melbourne, Australia
Jon.Whittle@data61.csiro.au

### Burak Turhan
University of Oulu
Oulu, Finland
burak.turhan@oulu.fi

## ABSTRACT

Human values such as *integrity, privacy, curiosity, security,* and *honesty* are guiding principles for what people consider important in life. Such human values may be violated by mobile software applications (apps), and the negative effects of such human value violations can be seen in various ways in society. In this work, we focus on the human value of *honesty*. We present a model to support the automatic identification of violations of the value of honesty from app reviews from an end-user perspective. Beyond the automatic detection of honesty violations by apps, we also aim to better understand different categories of honesty violations expressed by users in their app reviews. The result of our manual analysis of our honesty violations dataset shows that honesty violations can be characterised into ten categories: unfair cancellation and refund policies; false advertisements; delusive subscriptions; cheating systems; inaccurate information; unfair fees; no service; deletion of reviews; impersonation; and fraudulent-looking apps. Based on these results, we argue for a conscious effort in developing more honest software artefacts including mobile apps, and the promotion of honesty as a key value in software development practices. Furthermore, we discuss the role of app distribution platforms as enforcers of ethical systems supporting human values, and highlight some proposed next steps for human values in software engineering (SE) research.

## 1 INTRODUCTION

Human values such as *integrity, privacy, curiosity, security,* and *honesty,* are the guiding principles for what people consider important in life [11]. These values influence the choices, decisions, relationships, and the concept of ethics for people and society at large whether or not they are formally articulated in this terminology [58]. The relationship between human values and technologies is important, especially for ubiquitous technologies like mobile software applications (apps) [46]. Mobile apps are a convenience to modern society and have seen usage in carrying out both simple and complex tasks, from entertainment (e.g., video sharing apps) and health (e.g., fitness trackers) to finance (e.g., banking apps). End-users of these apps hold certain expectations influenced by their human values considerations, e.g., the privacy of data, transparency of processes in apps, and ethical behaviour of platforms and software companies [46]. The violation of these value considerations is detrimental to the end-user, software platforms, companies, and society in general [67].

Recent work on human values in software engineering (SE) based on the Schwartz theory of basic human values [58, 59] have mapped human values to specific ethical principles. For example, Perera et al. mapped values to the GDPR principles [52] and Winter et al. mapped values to the ACM Code of Ethics [69]. Other studies such as [46] have explored the violation of human values in mobile apps using app reviews as a proxy. The recent study by Obie et al. showed that the violation of *honesty* (a sub-item of *benevolence* based on Schwartz theory [58]) is violated by mobile apps [46].

Honesty, often perceived to be a very important human value [39], describes a character quality of being sincere, truthful, fair, and straightforward, and refraining from lying, cheating, deceit, and fraud [15]. The importance of the value of honesty is clearly articulated in the ACM Code of Ethics: "Honesty is an essential component of trust. A computing professional should be transparent and provide full disclosure of all pertinent system limitations and potential problems. Making deliberately false or misleading claims, fabricating or falsifying data, and other dishonest conduct are a violation of the Code..." [21]. Nonetheless, there have been many flagrant violations of the value of honesty by mobile app platforms and software companies [17, 23, 65].

Consider the following example of the violation of honesty. The dating platform (Match.com) has been accused of faking love interests using bots and fake profiles to fool consumers into buying subscriptions and exposing them to the risk of fraud and other deceptive practices [54]. During a period of over three years, the company allegedly delivered marketing emails (i.e., the *"You have caught his eye"* notification) to potential consumers after the company's internal system already flagged the message sender as a suspected bot or scammer. The company also violated the "Restore Online Shopper's Confidence Act" (ROSCA) by making the unsubscription process tedious; internal documents showed that users need to make more than six clicks to cancel their subscription, resulting in the U.S. Federal Trade Commission (FTC) suing Match.com for "deceptive advertising, billing, and cancellation practices" [54].

Consider the more recent example of Shaw Academy who offered users a free trial to its online education platform and charged them a subscription fee even after they had cancelled before the end of the trial period and refused to refund the users [70]. The outcome of an investigation by the Australian Competition & Consumer Commission (ACCC) ordered the company to refund approximately $50,000 to the affected users and pledge to improve their system [70]. Here is an example review of dubious charges to a user account for a calendar reminder app:

*"I've been charged $45+ on 2 separate occasions in the month I've had the 'premium' version. It advertises $3.50 for a premium subscription but saw nowhere where it said they would make additional charges. There is absolutely no reason a calender reminder app should charge this much without telling you or without being deceptive."*

Other examples include companies deliberately hiding data breaches from the authorities and customers [7, 61]. These violations of the value of honesty result in decreased trust from users, poor uptake of apps, and reputational and financial damage to the organisations involved. This also emphasises the need to consider human values more proactively in software engineering practice.

To detect the violation of the value of honesty in mobile apps, we utilised user's comments expressed in app reviews, as reviews are a valuable resource and have been shown to be a proxy for detecting users' challenges and requirements [5, 14, 22, 46, 62]. To this end, we formulated the identification of the violation of honesty in app reviews as a classification problem. We trained and compared five machine learning models based on a manually annotated dataset to learn the features that are representative of the violation of honesty in app reviews. The best performing model (Support Vector Machine) has an F1 score of 0.89, a precision of 0.94, and a recall of 0.84. Additionally, beyond the automatic detection of honesty violations, this work also aims to understand the different categories of honesty violations expressed in app reviews. Thus, we manually analysed reviews containing honesty violations. Our resulting taxonomy shows that honesty violations can be characterised into ten categories: **unfair cancellation and refund policies, false advertisements, delusive subscriptions, cheating systems, inaccurate information, unfair fees, no service, deletion of reviews, impersonation,** and **fraudulent-looking apps**. In summary, this work makes the following contributions:

- We present machine learning models and datasets to aid the automatic detection of the violation of the human value of honesty in reviews. Our publicly available replication package supports researchers and practitioners to adapt, replicate, and validate our study [6].

- We provide insight into the different categories of honesty violations prevalent in app reviews by creating a taxonomy based on a manual analysis of the honesty violations dataset.

- We present a set of practical recommendations and future research directions to deal with the challenges of the violations of the human value of honesty in apps that would benefit end-users and society.

## 2 RELATED WORK

### 2.1 Mining App Reviews

Several studies have been carried out to provide insights into user reviews and how these reviews can aid software professionals in app maintenance [9, 51, 60] and evolution [12, 35, 36, 49]. Guzman and Maalej adopted NLP techniques to locate fine-grained app features in reviews with the aim of supporting software requirements tasks [22]. A related work utilised Latent Dirichlet Allocation (LDA) technique and linguistic rules to group feature requests from users as expressed in their reviews, and the results from this study showed that users care about frequent updates, improved support, more customisation options, and new levels (for game apps) [25].

Some studies have focused on the automatic classification of app reviews into useful categories. To aid software professionals in prioritising accessibility issues, AlOmar et al. developed a machine learning model for identifying accessibility-related complaints in app reviews [5]. Panichella et al. introduced a taxonomy for classifying app reviews and using a combination of NLP and sentiment analysis classified app reviews into their proposed taxonomy [50].

Other works have introduced tools to support the extraction of insights from app reviews. For example, Vu et al. proposed MARK, a keyword-based tool for detecting trends and changes that relate to occurrences of serious issues in reviews [56]. Similarly, Di Sorbo et al. introduced SURF, a tool that condenses thousands of reviews into coherent summaries to support change requests and planning of software releases [14].

The above studies show that app reviews are a useful resource for gathering requirements, detecting issues, and more generally for supporting software professionals in evolving their apps. This work also aims to support app maintenance and evolution by effectively detecting potential violations of the value of honesty from the user's perspective in app reviews. In addition, it would aid software professionals in delivering software products that build trust in users, as the honesty (real or perceived) of companies can affect how users engage with their products [71].

### 2.2 Human Values in Software Engineering (SE)

Human values are enduring beliefs that a specific mode of conduct or end state of existence is personally or socially preferable to an opposite or converse mode of conduct or end state of existence [57]. Human values have been well-studied in the social sciences and have begun to see adoption in other fields including design [4] and software engineering (SE) [34, 41].

The study of human values in SE is a relatively nascent line of research [42, 53] and is mostly based on the widely accepted and adopted Schwartz theory of basic human values [58, 59]. The Schwartz theory is built on a survey conducted in over 80 countries covering different demographics. This theory categorises values into 10 broad categories, namely: *self-direction, stimulation, hedonism, achievement, power, security, conformity, tradition, benevolence,* and *universalism*. These 10 categories in turn are made up of 58 value items, e.g., the value category of **benevolence** covers the value items of **honesty**, *responsible, helpful, forgiving, loyal, mature love, a spiritual life, meaning in life,* and *true friendship* (c.f [58]). However, our focus in this work is on the value item of *honesty*, based upon the prevalence of the value category of *benevolence* in prior research [46], the recent cases of the violations of *honesty* by companies in the media, e.g., [54, 70], and the need to understand this phenomenon more closely in SE.

Studies in the social sciences have investigated the value of (dis)honesty at the individual and organisational levels [19], and the policy implication of dishonesty in everyday life [38]; while others have explored the motivation for dishonest behaviours [1] including students in classroom settings [30] and workers in crowd-working environments [26]. Keyes argues that euphemising the violation of the value of honesty desensitises people to its implications and consequences in society [28].

However, within the context of SE, Whittle et al. argued that software companies need to consider human values in the development of software systems and make them "first-class" entities throughout the software development life cycle [67]. Another study made a case for the evolution of current software practices and frameworks to embed human values in technology instead of a revolution of the SE field [24].

Another line of research considered methods for measuring human values in SE. For example, Winter et al. introduced the Values Q-sort instrument for measuring human values in SE [69]. Applying the Values Q-sort instrument to 12 software engineers resulted in 3 software engineer values "prototype". Similarly, Shams et al. utilised the portrait values questionnaire (PVQ) to elicit the values of 193 Bangladeshi female farmers in a mobile app development project [63]. The result of the study showed that conformity and security were the most important values while power, hedonism, and stimulation were the least important. More recently, Obie et al. argued that the instruments for eliciting and measuring values should be contextualised to specific domains [47].

Recent studies have adopted the use of app reviews as an auxiliary data source for eliciting values requirements. Shams et al. analysed 1,522 reviews from 29 agricultural mobile apps to understand the values that are both represented and missing from these apps [62]. Obie et al. proposed a keyword dictionary-based NLP classifier to detect the value categories violated in app reviews [46]. The results of the application of the classifier to 22,119 reviews showed that benevolence and self-direction were the most violated categories while conformity and tradition were the least violated.

The studies highlighted have been instrumental in pushing the frontiers of human values in SE, and the closely related works such as [46, 62] have provided insights to violations of value categories. Our work complements these by zooming in on a specific value item; **honesty** (within the most violated category of **benevolence** [46]), to provide a more nuanced understanding of its violations. In addition, we provide a taxonomy of the different categories of honesty violations in reviews to better understand how the violation of the value of honesty is reported. We hope that other researchers would be encouraged to investigate other specific value categories, and more generally explore the field of human values in SE.

## 3 RESEARCH DESIGN

Our goal in this study is to automatically identify reviews discussing honesty values and indicating from these reviews to determine the different types of honesty violations documented. To do this we define the following research questions (RQs):

**RQ1.** *Can we effectively identify reviews documenting honesty violations automatically?*

**RQ2.** *What types of honesty violations are reported in these app reviews?*

### 3.1 A Dataset of Honesty-related Reviews

The first step to answering our RQs is creating a dataset of user reviews documenting perceived honesty violations by apps.

*3.1.1 Data Collection.* To build this dataset, we collected a total of 236,660 reviews - 214,053 reviews from the public dataset of Eler et al. [18], and an additional 22,607 reviews from the public dataset of Obie et al. [46]. These reviews were collected from a total of 713 apps in 25 categories. The apps and reviews were intended to cover a diverse range of categories and audiences. Table 1 summarises the statistics of our combined app review dataset. Our dataset can be found here [6].

*3.1.2 Data Labeling.* Given the sheer size of the dataset and the manual labour required to label the dataset, we used two approaches to label the 236,660 reviews: a keyword-based approach and manual labeling. We first adopt a set of keywords to filter the 236,660 reviews to include those related only to the value of honesty. These keywords are based on the dictionary of human values created by Obie et al. [46]. The set of keywords comprise a total of 48 words semantically related to honesty. The keywords are available in [6]. After applying this keyword filter, the number of reviews was reduced from 236,660 reviews to 4,885 potential candidate honesty-related reviews (we call these 4,885 reviews **honesty_potential** reviews).

However, adopting a keyword-based approach is error-prone and may result in a lot of false positives. Hence, we manually analysed the **honesty_potential** reviews to exclude false positives. Moreover, the application of keywords filter and subsequent manual analysis check have been applied in recent studies [5, 18].

The **honesty_potential** reviews were labelled and validated in 25% increments in the following manner. The first analyst labelled the first 25% percent of the **honesty_potential** reviews to determine which of the reviews contain the violation of the value of honesty as perceived by the user in the review. The second analyst validated the outcome. The disagreements were resolved in a meeting using the negotiated agreement approach to address issues of reliability [8, 40]. Then the next 25% were labelled by the first analyst, validated by the second analyst, and disagreements resolved in a meeting as in the first round. The same procedure was repeated for the third and fourth rounds of the labelling process. Also, the labelling and validation were done over eight weeks to avoid fatigue. Based on our manual labelling, we found that out of the 4, 885 filtered reviews (the **honesty_potential** reviews), only 401 were honesty violations reviews, i.e., true positives. We refer to these 401 honesty violations reviews as **honesty_violations** reviews.

**Table 1: Statistics of the dataset.**

| | |
|---|---|
| Number of Apps | 713 |
| App Categories | 25 |
| All Reviews | 236,660 |
| Honesty-related Reviews (after keywords filter) | 4,885 |
| Honesty Violation Reviews (after manual validation) | 401 |

Next, we randomly selected 401 reviews from the remaining 4,484 **honesty_potential** reviews (4,885 **honesty_potential** reviews - 401 **honesty_violations** reviews). We refer to these 401 reviews, which contain honesty-related keywords (but not violations), as **honesty_non_violations** reviews. We used a total of 802 reviews: 401 **honesty_violations** and 401 **honesty_non_violations** reviews to build a balanced dataset called **honesty_discussion** dataset for training and evaluating machine learning models in Section 4. We note here that using the manually validated false-positive *honesty_non_violations* reviews is important for machine learning models. It is because these reviews include certain keywords syntactically related to honesty but semantically irrelevant to honesty violations - an important difference we want our models to learn. In summary, the **honesty_discussion** dataset consists of 802 reviews: 401 **honesty_violations** reviews and 401 **honesty_non_violations** reviews. Other studies have used similar numbers of text documents in classification tasks [32, 33].

## 4 AUTOMATIC CLASSIFICATION OF HONESTY VIOLATIONS (RQ1)

### 4.1 Approach

Manually classifying honesty violations in app reviews is challenging for practitioners because it is error-prone, labor-intensive, and demands substantial domain expertise. Hence, an automated approach is required to recognise honesty violations in app reviews. This research question aims to develop machine learning models to differentiate between honesty and non-honesty reviews automatically. The machine learning models are applied on the 802 **honesty_discussion** dataset which consists of 401 **honesty_violations** reviews and 401 **honesty_non_violations** reviews.

*4.1.1 Data Preparation.* We applied some common techniques to remove possible noise from the **honesty_discussion** dataset. This step was needed so a learning model can classify reviews correctly. To achieve this, we applied natural language processing techniques such as removing capitalisation, removing emojis, tokenising, removing stop words, and removing punctuation.

**Case Normalisation:** is the process of transforming original review texts into their lower case. This type of text cleansing helps us avoid repeated features of the same words with different font cases (e.g., "Honesty" and "honesty"). Furthermore, converting the text into its lower case does not affect its context as well as the users' expressions in our scenario.

**Emoji Removal:** Emojis are icons or a few Unicode characters that allow users to convey ideas, concepts, and emotions. If emojis are not carefully preprocessed, they can potentially affect the performance of a model in terms of accuracy. Hence, we removed Emoji from the review texts.

**Tokenisation:** is the process of splitting each original text into a set of words that do not contain white space. We divided apps reviews into their constituent set of words.

**Stop-Word Removal:** Stop words such as *is, am, are, for, the,* and others do not contain conceptual meaning of a review and create noise for a classification model. Removing stop words from the review texts helps us avoid repeated features of the same phases (e.g., "the bank account" and "bank account"). In our experiment, we used a comprehensive set of stop words that are well-known to the natural language processing community[1].

**Punctuation Removal:** We observed many reviews in the data collection containing punctuation such as *"..., ??, :(,"* and others that do not significantly contribute to a classification model. Hence, we removed punctuation from the app reviews.

*4.1.2 Feature Extraction.* After cleansing and preprocessing the dataset, we converted the app reviews in the dataset into their vector representation by using the pre-trained Bidirectional Encoder Representations from Transformers model [13], so-called BERT[2]. This is a language representation model trained on the BooksCorpus with 800 million words [72] and English Wikipedia with 2.5 billion words. The model receives a sequence of words as input and outputs a sequence of vectors. The model converted the review texts with different words into 768-dimensional vectors used as an input in a machine learning model. Each of these vectors is estimated by the average of embedded vectors of its constituent words. For instance, given a review text $s$ that consists of $n$-words, $s = (w_1, \ldots, w_n)$, then, $\vec{s} \approx \frac{1}{n}(\vec{w_1} + \ldots + \vec{w_n})$, where $(\vec{w_1} + \ldots + \vec{w_n})$ are the embedded vectors of $(w_1, \ldots, w_n)$. Furthermore, these vectors capture both a semantic meaning and a contextualised meaning of their corresponding app reviews.

*4.1.3 Model Selection and Tuning.* Selecting a classification model that yields the optimal result is challenging. We selected five models, such as Support Vector Machine (SVM), Decision Trees (DT), Neural Network (NN), Logistic Regression (LR) and Gradient Boosting Tress (GBT) that are commonly used for text classification in the natural language processing community [2]. Below is a brief description of each classification model used in our work.

**Logistic Regression (LR)** is a linear classifier. The data is fitted into a logistic function that generates the binary output such as 0 (i.e., an honesty_non_violation app review) or 1 (i.e., an honesty violation app review) based on probability.

**Support Vector Machine (SVM)** [44] is a classifier that finds hyperplane(s) in N-dimensional space (i.e., the number of features), which can further distinguish the data into multiple categories.

**Decision Trees (DT)** is one of the ensemble learners that builds trees for classification. Each tree represents a particular characteristic of the data. Given a 768-dimensional vector representation of a particular review text, DT classifies the review text into the category selected by most trees.

---

[1]The stop words can be accessed at https://gist.github.com/sebleier/554280#gistcomment-3126707
[2]The pre-trained BERT uncased model can be downloaded at https://huggingface.co/bert-base-uncased.

**Table 2: Comparison of confusion matrix and Matthews correlation coefficient (MCC) of classification models.**

|                | SVM   | LR    | NN    | RF    | GBT   |
|----------------|-------|-------|-------|-------|-------|
| True negative  | 0.432 | 0.407 | 0.358 | 0.371 | 0.358 |
| True positive  | 0.457 | 0.469 | 0.482 | 0.420 | 0.420 |
| False positive | 0.025 | 0.049 | 0.099 | 0.085 | 0.099 |
| False negative | 0.086 | 0.074 | 0.062 | 0.124 | 0.124 |
| MCC            | 0.785 | 0.753 | 0.676 | 0.581 | 0.555 |

**Table 3: Comparison of classification models.**

|           | SVM   | LR    | NN    | RF    | GBT   |
|-----------|-------|-------|-------|-------|-------|
| Accuracy  | 0.889 | 0.877 | 0.840 | 0.790 | 0.778 |
| Precision | 0.949 | 0.905 | 0.840 | 0.829 | 0.810 |
| Recall    | 0.841 | 0.864 | 0.886 | 0.773 | 0.773 |
| F1 score  | 0.892 | 0.884 | 0.857 | 0.800 | 0.791 |

**Gradient Boosting Trees (GBT)** is one of the ensemble learners that builds trees and boosts them for classification. When a new tree is created, it corrects errors of previous trees fitted on the same provided data. This repeatedly correcting errors process is known as the boosting process. In addition, the gradient descent algorithm is used for optimisation during the boosting process. Thus, the method is called gradient boosting trees. The model classifies app reviews into a category based on the entire ensemble of trees.

**Neural Network (NN)** is a multilayer perceptron model which contains a set of interconnected layers where each layer contains a finite number of nodes. Each neural network architecture has one input layer, at least one hidden layer, and one output layer. The input data is transformed layer by layer via the activation function(s). During the training process, optimisation techniques such as stochastic gradient descent are used to optimise the performance of the model. The classified category of a particular app review is the collected result from the output layer.

Finding the hyperparameters for models to generate the optimal results is known as the fine-tuning process. We use grid search cross-validation to perform an exhaustive search to find the best set of hyperparameters for each classifier. To reproduce our results, we provide the selected hyperparameters for each selected model and the open-source GitHub repository in [6].

*4.1.4 Cross Validation.* To estimate the variance of the performance for each classification model, we used a 10-fold cross-validation technique. Here, we split the dataset in Section 3.1 into 10 chunks of data that contains an equal number of app reviews. Then, we perform the evaluation process where the training dataset contains 9 chunks of data, and another chunk of data is used as the testing dataset. Note that this is repeated until each chunk of data has been used as the testing dataset once. This approach helps us understand how well our selected models perform on unseen data.

## 4.2 Results

In this section, we report the results of our experiment evaluating the performance of the different machine learning models. We adopted the generally accepted metrics of ***accuracy, precision, recall***, and ***F1 score*** for this purpose. Other metrics such as the Matthews Correlation Coefficient (MCC) and confusion table are shown in Table 2.

We note here that all of the models performed well (with F1 scores of 0.79 and above).

Table 3 shows the results of 5 different machine learning classification algorithms. The SVM algorithm came out to be the best performing model with an accuracy of 0.88, precision of 0.94, recall of 0.84, and an F1 score of 0.89. The second-best performing algorithm is the LR model, with an accuracy of 0.87, precision of 0.9, recall of 0.86, and an F1 score of 0.88.

Furthermore, the high performance of our SVM model makes it useful in practical applications for detecting the violation of the value of honesty in reviews.

*4.2.1 Comparison with Baselines.* One of the aims of our work is to introduce an automatic method for detecting honesty violations reviews that performs better than current approaches. Similar studies on text classification have compared their approaches to either the current state-of-the-art or a baseline random classifier [5, 37]. Hence we compare our best-performing machine learning model (SVM) with a baseline random classifier only since there is no current state-of-the-art in detecting the violation of honesty in app reviews, similar to what recent works have done [5, 37].

We used the statistics of our dataset to compute the metrics of the random classifier. The precision of a random classifier can be computed by dividing the number of honesty violation reviews by the total number of reviews:

$$precision = \frac{401}{236,660} = 0.0017$$

The recall is 0.5, as there are only two outcomes for a review classification: honesty violations reviews or honesty_non_violations reviews, with a 0.5 probability of a review containing the violation of the value of honesty. Based on the precision and recall values, we compute the F1 score of the baseline random classifier as:

$$F1\ score = 2 * \frac{0.0017 * 0.5}{0.0017 + 0.5} = 0.0034$$

Table 4 summarises the comparison of our best-performing machine learning model (SVM) with the baseline. As can be seen, the SVM model has a better performance than the baseline random classifier. Our SVM model has an F1 score of 0.89, while the baseline random classifier has F1 score of 0.0034, respectively. Table 4 also shows that our SVM model surpasses the baseline random classifier by 262.353 times in detecting honesty violations reviews.

> **RQ1 Answer**: The SVM model surpasses the baseline random classifier in identifying the violation of the value of honesty in reviews. Our model achieves an F1 score of 0.892 with an improvement of 262.353 times the baseline random classifier in classifying honesty violation reviews from honesty_non_violation reviews.

**Table 4: Comparison of our model to a baseline classifier.**

|  | Our (SVM) approach | | | Random classifier | | |
|---|---|---|---|---|---|---|
|  | Precision | Recall | F1 | Precision | Recall | F1 |
| Classification | 0.949 | 0.841 | 0.892 | 0.0017 | 0.5 | 0.0034 |
| Improvement | - | - | - | 558.235x | 1.682x | 262.353x |

## 5 CATEGORIES OF HONESTY VIOLATIONS (RQ2)

### 5.1 Approach

While the machine learning models in Section 4 could effectively distinguish between honesty violations reviews and honesty non-violations reviews, we are also interested in understanding the types of honesty violations reported in reviews. To this end, we applied the open coding procedure [20] on the 401 **honesty_violations** reviews. As discussed in Section 3.1, these reviews include honesty violations. First, an analyst followed the open coding technique to label all these 401 reviews and identified 10 types of honesty violations. The 401 **honesty_violations** reviews were assigned to these 10 categories. The results of the open coding were stored in an Excel spreadsheet file and shared with the second and third analysts. Then, the second analyst cross-checked the first 100 labelled reviews while the third analyst cross-checked the remaining 301 labelled reviews. Next, the first analyst held Zoom meetings with the second and third analysts to discuss and resolve the conflicts and disagreements. Note that the disagreements were resolved using the negotiated agreement approach [8, 40].

### 5.2 Results

Our analysis of the 401 **honesty_violations** reviews revealed 10 categories of honesty violations reported in app reviews. Below we provide a definition of these categories, sample reviews, and a summary of their prevalence. While we highlight the different categories within the violation of the value of honesty and provide example reviews, we note that the categories are not mutually exclusive. Table 5 shows these categories and the frequency of the corresponding reviews per category.

*5.2.1 Unfair cancellation and refund policies.* This category covers all reviews where the users perceive the cancellation and refund policy as unfair, nontransparent, or deliberately misleading. It also includes situations where the user feels that the developers deliberately make it difficult for the user to cancel their subscription. For example, in some apps, the user can sign up for a subscription with the click of a button within the app but cannot cancel the subscription from within the app; the user is asked to log in to a website to cancel the subscription. In other cases, the cancellation instruction is not clear and leads to a loop of cancellation steps. Examples of reviews claiming these practices include:

> "*The app allows you to accidentally sign up to premium with a push of a button. When you want to cancel, however, you can't do that via the app... You have to go to the webpage, enter details and cancel there.*"

> "*Deceptive billing practices - information on cancelling is circular; emailed a link that advises to email. [It] doesn't have colour tag*

*functionality across web and app; very poor UX and worse customer service.*"

Sometimes, the app also makes it easy for the user to mistakenly activate a premium subscription in the way the interface and flow are designed, e.g.:

> "*Use with caution. It's unscrupulous about signing you up for a subscription when you're skipping past the in-app ads. It's not made clear once you've subscribed, and there's no way of cancelling it through the app.*"

Another aspect of this category focuses on situations where the user perceives the refund steps and policies to be dishonest and unfair. This also involves situations where the refund policy does not cater to accidental subscriptions, e.g.:

> "*DO NOT SIGN UP FOR FREE TRIAL! IT IS A SCAM. YOU WILL GET CHARGED ANYWAY, AND YOU WILL NEVER GET YOUR MONEY BACK!! Once again, after numerous attempts to blame Google, this developer has still not refunded my $38. Once again, I cancelled 3 full days before the free trial ended but was still charged. Once again, [I] contacted the developer, who told me that I would receive a full refund within 7 to 10 days, and still nothing. I have saved the email, pricing this to be true. DO NOT TRUST THIS DEVELOPER. SCAM!!!!*"

*5.2.2 False advertisements.* This category relates to situations where the user perceives that the advertised features and functionalities of the app as described by the developers are not contained in the app. The user downloads the app or pays for a subscription on the basis of accessing certain functionalities or features only to find out the descriptions, including screenshots on the app distribution platform is different from the actual functionalities available in the app. Two examples of these are shown below:

> "*Couldn't find Google Assistant integration anywhere. Even though it's been advertised everywhere when searching the web for the app... It's even in the description of the app here. That's false advertising. I will edit my review when it's out of Beta and working in the final version.*"

> "*The app doesn't listen to the watch at all. I've tried completing and snoozing and it does nothing. The watch can only add tasks, so the screenshots they're sharing here are DECEPTIVE.*"

In some cases, the app lures users into downloading the app on the basis that it is free-for-use only for the user to find out that the free-for-use is a trial version for a specific time period and not perpetually free as implied in the app description:

> "*The actual free version doesn't allow you anything, not even to learn how to use the app properly. That role is filled by 7 days of free premium. The free, on the description, is a lie. Is a paid-only app with temporary free access to its full features that gets practically useless after the 7-day trial... I don't like to be lied to.*"

In addition, the app developers (through the app description) make promises to users to give them certain benefits like a free premium subscription when a particular action is carried out (e.g., inviting a particular number of friends to sign up). However, they never truly fulfil their promises when the user fulfils their end of the bargain. These unfulfilled obligations are perceived by the end-user as a violation of honesty, e.g.:

**Table 5: Frequency ($f$) of app reviews in the honesty violation categories (out of 401 total *honesty_violations* reviews – note that some reviews fall into multiple categories).**

| | Unfair cancellation and refund policies | False advertisements | Delusive subscriptions | Cheating systems | Inaccurate information | Unfair fees | No service | Deletion of reviews | Impersonation | Fraudulent-looking apps |
|---|---|---|---|---|---|---|---|---|---|---|
| $f$ | 48 (12%) | 55 (14%) | 33 (8%) | 93 (23%) | 15 (4%) | 106 (26%) | 64 (16%) | 6 (1.5%) | 9 (2%) | 29 (7%) |

🗩 *"I love this app however I sent the link to several friends and they got the app and I received no premium time whatsoever. Don't be dishonest with your apps. That's lame."*

Another example relates to scenarios where the user is invited to make certain commitments based on a future reward and the developers bail out on their prior commitment:

🗩 *"Shame on Them! Liars. I paid for the season pass TWICE (ONCE for my apple device and the other for my Samsung Device). I was falsely promised access to ALL FUTURE CONTENT. Now they are trying to charge me for the Parisian Inspired TOKENS! HOW DARE THEY LIE AND BAIT AND SWITCH."*

*5.2.3 Delusive Subscriptions.* Any review describing complaints related to unfair or nontransparent automatic subscription processes is classified under this category. There are instances where no notifications are provided to let the user know they are subscribed to the app or premium version of the app, and the user only finds out about the subscription from the deductions in their bank accounts:

🗩 *"I just realised that I have been charged for some crappy premium service fee which I had no idea about when using the app. Why is this charge by default? Why was I not informed in the first place? Beware of scam for useless monthly premium fees!"*

🗩 *"I can't believe I was charged 55.99. What are you giving me? Gold? I unsubscribed but saw mysterious charge in my bank account."*

Additionally, there is the issue of lack of user consent in the subscription process where certain apps do not provide a confirmation mechanism that prevents accidental subscriptions by the user, e.g.:

🗩 *"Made me pay 1 year worth of subscription without my confirmation. Only used its free trial because I had to use it once. What a scam..."*

In some scenarios, the automatic subscription is hidden behind an in-app ad/feature, and an unsuspecting user who clicks on the feature is automatically subscribed to the premium version of the app without a clear warning or confirmation, e.g.:

🗩 *"Deceptive practices. If you click the in-app "ad" that simply says enable notifications, you'll automatically be signed up and billed for their premium service. This bypasses the Google/Apple stores subscription model and bills your card directly. Not to mention it's impossible to downgrade from this service in the app itself; you have to visit their website, which is a deliberately obstructive hurdle considering you can upgrade in the app just fine."*

*5.2.4 Cheating systems.* All reviews concerning the user's perception of fraud by other persons or cheating within the inner workings of the app are classified under this category. Users complain of unfairness in either the process or outcome of the app, especially processes/outcomes that are supposedly statistically random. While accusations of this kind from the users are prevalent and subjective, they may not realistically be the case. However, we labelled these kinds of reviews based on the *perception* of the users as captured in their comments. Reviews related to this category are mostly found in games or game-like systems. For example:

🗩 *"This game cheats. It uses words not found in the dictionary. Also it told me a word was unplayable, but it was the first best word option."*

🗩 *"I play it with my sister often. However, there is the problem of the game and AI cheating. I rolled a 2 and a 3 at the start of the game and it moved me FOUR spaces forward not five. Four. That happened several times and I can assure you I was looking everytime it happened. I am very disappointed at the fact this game is cheating..."*

In some of the reviews, users complain that the game works properly when the user loses and parts with money and only freezes when the AI system in the app is about to lose. Based on the reviews, the users seem to be using real money in the games/apps. This complaint is a recurring theme within this category:

🗩 *"You have to pay for it, then the game just freezes when you win against the CPU? Reset it over and again, keeps freezing unless it rolls something to not land on my property. Also, is the dice rigged against the CPU? Honesty? With as much as I owned in the beginning, none of the 3 CPUs would land on anything I owned. Anytime the last CPU needs to raise money, game freezes, guess ya just can't win."*

🗩 *"there's a glitch in it that freezes the game from continuing when you're winning. The dice just disappears, but the trains and clouds and aircrafts keep moving. It's like It is designed so that one doesn't win them."*

🗩 *"When playing against the computers when you're about to win and bankrupt the final computer the game conveniently freezes. It does not allow you to win. Not a very fun game to play, I want my money back."*

We consider this category important as some of these apps require the use of real money to play or for in-app purchases. If apps are dishonest in the underlying process of the systems that are expected to be fair, then that constitutes not only the violation of the value of honesty, it might potentially be a crime. This is worth considering, especially when the exact issue is raised by several users:

🗩 *"Although you say that the dice is random, i cannot help but feel that it is rigged. Take a look at your reviews, there are many other players that feel the same. Can't be all of us are wrong. Or maybe we are suffering from mass hysteria?"*

Other non-games examples include cases where the user reports not having the full value of the fee they were charged for the app and feels cheated. For instance:

🗩 *"Whenever I pay for parking the app always steals 5 minutes off my parking time. For example, I pay for 60 minutes and the timer starts at 54 minutes and 59 seconds. I am very upset, this has been happening for a while and probably to many more people as well. That is a lot of money!"*

🗩 *"This app will not give you're requested amount of parking time. If you park for 15 minutes it will immediately say you have 11 minutes left. I understand that you have to charge but at least give me the requested amount of parking time."*

*5.2.5 Inaccurate information.* This category covers where users perceive that the app provides false or inaccurate information as captured in their reviews. This includes situations where inaccurate information can increase the likelihood of the user inadvertently making wrong selections at a cost to them. In the review below, the user complains about the design of an app feature tricks them into paying for the wrong parking spot:

🗩 *"When you need to pay for additional time, and click 'Recent' to pay for the most Recently parked in place - the first item is not the place you just parked in so it tricks you into paying for the wrong place (dark pattern). Please make the Recent accurately reflect the most recently parked in place."*

Another example review in this category is quite severe as it relates to a health emergency app providing potentially inaccurate information that might be detrimental to the user:

🗩 *"Try to use this in an actual emergency and you'll just end up as a dead idiot holding a cellphone. The information is either useless or completely false in most cases. Don't bother downloading."*

Other less severe but important reviews where the user perceives the app provides inaccurate information or notification are shown below:

🗩 *"Do not buy unless you are sure you want to. You will NOT be able to get it set up and working within the 15 minute refund window. The instructions online are so cryptic it (and wrong)."*

🗩 *"Very annoying every time when you open the app it shows you have a notification. Then checking your notifications you don't have any."*

*5.2.6 Unfair fees.* This category relates to issues surrounding what the user considers to be unfair fees or charges. This also applies to cases where the user feels that they have not received a fair deal or that the app charges more money than it ought to. Because the definition of honesty also covers fairness, we also consider these kinds of issues a potential violation of the value of honesty. In the example below, the user complains of being charged more than they think is fair; they were charged a car parking rate for parking a bike.

🗩 *"Went through the sign up process and parked my bike in a bike parking zone. Put in the correct zone details for the bike parking area and got charged a car parking rate. Rang support and they said there is no bike parking at that location. I explained there was and they told me to ring the council."*

Other examples of fees considered by the user to be unfair are:

🗩 *"The app charges you 0.25 per transaction. So I paid 0.75 to pay for parking it charged me 0.25 service fee then I extended my parking 0.25 and it charged me again 0.25!!! Biggest scam in the world."*

🗩 *"The only annoying things are that I have to buy any extra Monopoly Board in the same game when I already paid the main game. Can you not give extra Monopoly Boards in the same game for free. You are not fair!"*

This category is also reflected in the form of hidden charges where the user is not aware of subsequent charges made to their account. These hidden charges can take the form of a vague bill (as shown in the review below) or not notifying the user with respect to extra charges.

🗩 *"This is a notorious company with horrible app I've ever used. They hide the history and details very deep for you to check and trace. And the monthly bill is also vague. I experienced they secretly bill me!"*

🗩 *"LOOK OUT PEOPLE. THIS IS A SCAM. THEY DID NOT WARN OF A DEPOSIT FEE AND THEY TOOK 33% OF THE DEPOSIT. I RECOMMEND SUING THEM NOW."*

Another related issue within this category is dubious charges where the user account has been charged, and it is not clear why those charges occur. Abnormally high fees (more than the standard subscription fees) and overcharging of the user account are also captured under this category. For example:

🗩 *"It charged me £74.50 when I bought a ticket for £1.50 it's a absolute scam I want my money back!"*

*5.2.7 No service.* This category mainly covers reviews in which the user complains of not being able to access the app's main functionality after purchase, leading to undesirable consequences for the user. The main difference between the *false advertisement* category and this category is that the former deals with features/functionalities of the app that do not work as advertised. The latter deals with situations where the app does not work at all, i.e., does not even serve its main purpose for the user after the user has made financial commitments in the form of a purchase or subscription. In the example below, the user is fined for illegal parking after paying for parking using the app:

🗩 *"Horrible experience with this app. Causing a lot of frustrations with users. when it fails and I get a ticket there is no much help I can get. sometimes I just pay the fines just because the complaint system is awfully inconvenient. I feel cheated and it looks like a money making tool for whoever is collecting the fines."*

Another related example is shown below:

🗩 *"I spent 20 euros with all the DLCs included, I feel pretty deceived not being able to play the game."*

*5.2.8 Deletion of reviews.* This category highlights reviews where the app developers are suspected of deleting reviews left by the user, especially negative reviews. A review captures user feedback, describing their experience of an app, and intending users of an app typically consult the reviews left by other users on the app distribution platform before downloading the app [46]. Thus, the act of deleting unfavourable reviews by the app developers is perceived as a dishonest practice by the users because leaving only positive reviews may not paint an accurate picture of the app. Users may also feel like the app developers are trying to hide their complaints or other nefarious practices.

It can be argued that certain comments are deleted by app developers because those comments contain ad hominem attacks from the users instead of complaints relating to the app itself. While it is debatable whether app developers are justified in deleting perhaps vitriolic ad hominem comments, we do not make any judgement as to this but simply categorise users' perceptions and complaints of this practice as captured in their reviews. Examples of reviews depicting this accusation are shown below:

> 🗨 *"I left them a negative review and the developer deleted it. Now I'm going to review them on YouTube and all social media platforms. Basically, they are scammers."*

> 🗨 *"Deleted my honest review. Warning. Steer clear. They keep trying to make you slip up and pay for premium. I signed up for a free trial last year and they make it too difficult for you to find where to cancel. Was charged about $40... shame such a good app is tarnished by such shady practices."*

*5.2.9 Impersonation.* An impersonation is an act of pretending to be another person or entity [16]. It also involves the act of giving a false account of the nature of something. This category covers all reviews relating to impersonation or misrepresentation by the app or app developers. This includes scenarios where an app pretends to have the authority of (or relationship to) an organisation when in reality, it has no such relationship. An example review is captured below:

> 🗨 *"STAY AWAY... this app is a scam. the stickers make it look like it's Brisbane council approved. it's not and they are no help. I still got a fine for using the app correctly and the Brisbane council parking police have no access to check if you have paid or not and do not accept this as a payment method."*

Another example in this category reflects situations where users feel that they are interacting with bots instead of humans when they have signed up to the platform to interact with humans. This is similar to false advertising-related lawsuits of the Match.com platform described in section 1. An example of this is:

> 🗨 *"Good game, fake players online. I wanted a challenging Monopoly game. But when I start. I can tell that some are bots not real people online. For example, they quickly trade when it is their turn. A normal human will take some time to choose options."*

*5.2.10 Fraudulent-looking apps.* This category includes reviews reporting suspicious-looking apps based on observations of users or apps deemed to be fake by the users. We created a separate category for these kinds of reviews. Although the users flag the apps in these reviews as fraudulent, they do not provide specific reasons for their accusations beyond their perception of the app as fake or fraudulent. Furthermore, these types of reviews do not fit any of the categories described above, and we sought to highlight them based on the user accusations captured in their reviews. Examples of these reviews include:

> 🗨 *"...Be careful with this kind of dishonest apps"*

> 🗨 *"This is a fraud app don't download"*

---

> **RQ2 Answer:** The result of our analysis of the honesty violations dataset shows that honesty violations can be characterised into ten categories: unfair cancellation and refund policies, false advertisements, delusive subscriptions, cheating systems, inaccurate information, unfair fees, no service, deletion of reviews, impersonation, and fraudulent-looking apps.

## 6 DISCUSSION AND RECOMMENDATIONS

### 6.1 Technology (Mobile Apps) as Values Artefacts

Software artefacts such as mobile apps, like other technological artefacts express human values [66]. Although less well articulated, human values may be reflected throughout the different phases of the software development life cycle [45]. Values are represented in the conception and abstraction of ideas, in the way software features are arranged, described and even implemented and these embodied values are typically those of their creators, e.g., software developers and other stakeholders [31].

Some studies have argued that technological artefacts are value-agnostic tools that can be used for good or bad (i.e., theory of social determination of technology) [27], while others contend that technological artefacts are not value-agnostic, i.e., they hold value qualities and promote certain values over others [68], e.g., the bitcoin blockchain technology [43] is an embodiment of the value category of self-direction. Irrespective of the sociotechnological stance on values in technological (software) artefacts, there is an agreement on the role of software artefacts in changing habits in people and influences society in general, despite the intentions of the software companies behind these artefacts [3, 46]. Sullins writes, "Since the very design capabilities of information technology influence the lives of their users, the moral commitments of the designers of these technologies may dictate the course society will take and our commitments to certain moral values will then be determined by technologists" [64].

Furthermore, while we do not conflate values with ethics (values are the guiding principles of what people consider in life [57] while ethics are the moral expectations that a society agree upon to decide which values are acceptable or not [66]), the value of honesty is an ethically desired value in most societies. Thus we argue for a conscious effort in developing honest software artefacts including mobile apps, and the promotion of honesty in software development practices. Our intention in this paper is not to serve as moral arbiters of values in mobile apps (or other software artefacts) but rather to promote a healthy discussion of these issues in the software research and development community, and point the field towards a critical technical practice of mobile SE, i.e., the reflective work of sociocultural criticisms, highlighting the hidden assumptions in technical processes, and the interaction between the social, cultural and technical aspects of (mobile) SE.

### 6.2 The Role of App Distribution Platforms

App distribution platforms such as the Apple store and the Google Play store have an important role to play in supporting the values and minimising their violations in apps published on their platforms. They can serve as enforcers of ethical systems supporting values such as honesty, akin to the manner in which they protect end-users' devices from malicious apps. For instance, they can ensure that app developers are transparent in their billing process and enforce a mandatory multi-step (at least two steps) confirmation not only for subscription but also for in-app purchases.

Another issue on the violation of the value of honesty is related to the non-transparency in the subscription process in apps. For example, while some apps provide a reminder to the user before the end of a trial period so the user can decide to cancel their subscription or progress to a premium service, some other apps provide no reminder whatsoever. A reminder-to-cancel (or upgrade) feature for apps can be necessitated by the distribution platforms to protect the end-user from unintentional subscriptions.

In addition, for games or game-like apps involving the use of money for play, end-users perceptions of unfairness in these systems can be assuaged by a practice of auditing the systems to ensure statistical outcomes that are not only probable but fair to both the end-user and app developers alike, similarly to the way casino systems are routinely audited for fairness and transparency. The results of the audits can then be shown as part of the app information on the app stores.

## 6.3 Transparent Policies and Agreements

In cases of disputes between end-users and app vendors, where an end-user perceives that they have been unfairly treated, it is typical for the app vendors to refer the end-user to the end-user licence agreement (EULA) signed by the end-user during installation [29]. An EULA is a legally binding contract between the end-user and the app vendor [10].

Some app vendors place their data handling and billing processes in the fine prints of EULAs that are typically difficult to understand by the average user because they are written in legal terms [29]. Some studies have also shown that most end-users who clicked "I agree" do not understand the terms to which they agreed and often expressed genuine concern when the terms are expressed to them [10]. Thus it is important to develop transparent legal policies and easy-to-comprehend EULAs to inform and empower the end-user, and help them understand the terms and implications of these kinds of legal contracts. Transparency and comprehensibility would alleviate wariness and misgivings in this area. Also, we reiterate the position of O'Neill [48], that while transparency may undo secrecy, "it may not limit the deception and deliberate misinformation that undermine relations of trust. If we want to restore trust we need to reduce deception and lies, rather than secrecy" [48]. This area is particularly ripe for interdisciplinary research between the computing sciences, humanities, and law.

## 6.4 Human Values in SE Research

Research in the broader area of human values in SE is still in its early stages [53]. While the investigation of well-known values such as privacy and security have been considerably developed, other values such as honesty, curiosity, independence have received little attention, possibly due to the subjective and abstract nature of these concepts. This and other recent related works are based on an adaptation of the Schwartz theory of basic human values [58]. However, the nascent field of human values in SE may benefit from new conceptual theories of human values that are more situated closely within SE.

Furthermore, there is the need for the development of tools and techniques, not only in detecting the violation of human values in software artefacts but also providing automatic recommendations for possible fixes. Directions for future work may include the following: the development of approaches for generating end-user comprehensible EULA templates supporting values, approaches for evaluating and auditing fairness in games and game-like systems to support statistically probable results, and modules for static and dynamic analysis tools to detect specific values defects. Another area worth investigating is the development of tools for supporting the inclusion of values throughout the software development lifecycle and the resulting software artefacts including mobile apps.

## 7 THREATS TO VALIDITY

This section outlines the possible limitations and threats to the validity of our study.

**Internal Validity.** The qualitative process of building the **honesty_discussion** dataset in Section 3.1 and categorising the different types of honesty violations in Section 5.1 might be biased and error-prone. Hence, it might have introduced some threats to the internal validity of the study. We used three techniques to mitigate such threats. First, the qualitative analysis was conducted iteratively over an ample timeframe to avoid fatigue. Second, each review was analyzed by one analyst and validated by at least one other analyst, followed by several meetings between the analysts to resolve any disagreements and conflicts. Third, the analysts have extensive research experience in the area of human values.

**Construct Validity.** The analysts might have had different interpretations on the definition of the value of honesty. Our strategy to minimise this threat was making sure the analysts carefully examined seminal papers [58, 59] on the Schwartz theory, formal definition of honesty from dictionaries, and existing software engineering research on human values, including honesty [46, 62]. In this study, among many options, we used five machine learning algorithms to detect honesty violations reviews and four metrics to evaluate the algorithms. Peters et al. [55] claim that it is impracticable to use all algorithms in one study. Hence, we accept that applying other machine learning algorithms to our dataset may lead to different performances. The metrics precision, recall, accuracy, and F1-score used in this study are widely applied and suggested to evaluate machine learning models in software engineering.

**External Validity.** Our initial sample of app reviews was 236,660 reviews collected from [18] and [46], which was further reduced to 4,885 honesty-related reviews after applying the keywords filter. Our keyword filter may have introduced false negatives and potentially excluded honesty violations in the larger dataset. Hence, we cannot claim that our results are generalisable to all app reviews in the Apple App Store and Google Play Store and other platforms (e.g., online marketplaces).

## 8 CONCLUSION

Mobile software applications (apps) are very widely used and applied and hence need to reflect critical human value considerations such as curiosity, freedom, tradition, and honesty. The support for – or violation of – these critical human values in mobile apps have been shown to be captured in app reviews. In this work we focused on the value of honesty. We presented an approach for automatically finding app reviews that reveal the violation of the human value of honesty from an end-user perspective. In developing our automated approach, we evaluated five different algorithms using a manually annotated and validated dataset of app reviews. Our evaluation shows that the Support Vector Machine (SVM) algorithm provides a higher accuracy than the other algorithms in detecting the violation of the value of honesty in app reviews, and also surpasses a baseline random classifier with an F1 score of 0.89. We also characterised the different kinds of honesty violations reflected in app reviews. Our manual qualitative analysis of the reviews containing honesty violations resulted in ten categories: unfair cancellation and refund policies, false advertisements, delusive subscriptions, cheating systems, inaccurate information, unfair fees, no service, deletion of reviews, impersonation, and fraudulent-looking apps. The results of our study highlight the importance of considering software artefacts, such as mobile apps, as an embodiment of human values with consequences on end-users and society as a whole. We emphasise the role of app distribution platforms in supporting human values, such as honesty, on their platforms, and discuss the need for the software engineering research community to investigate methods and tools to better minimise the violation of human values in software artefacts.

## ACKNOWLEDGMENTS

This work is supported by ARC Discovery Grant DP200100020. Grundy is supported by ARC Laureate Fellowship FL190100035. Li is supported by ARC DECRA DE200100016.